\PassOptionsToPackage{table}{xcolor}
\documentclass[sigconf]{acmart}
\AtBeginDocument{%
  }
\copyrightyear{2026}
\acmYear{2026}
\setcopyright{cc}
\setcctype{by}
\acmConference[C\&C '26]{Creativity and Cognition}{July 13--16, 2026}{London, United Kingdom}
\acmBooktitle{Creativity and Cognition (C\&C '26), July 13--16, 2026, London, United Kingdom}
\acmDOI{10.1145/3803784.3807549}
\acmISBN{979-8-4007-2583-8/2026/07}

\newtoggle{comments}
\toggletrue{comments}

\iftoggle{comments} {
 \newcommand {\todo}[1]{{\color{red}\bf{TODO: #1}\normalfont}}
 \newcommand {\sarah}[1]{{\color{blue}\bf{ss: #1}\normalfont}}
}{
  \newcommand {\todo}[1]{}
    \newcommand {\sarah}[1]{}
}

    \newenvironment{myquote}{\begin{quote}\leftskip-14pt\rightskip-14pt}{\end{quote}}
        \newcommand*{\participant}[1]{{\small{\fontfamily{cmss}\selectfont{(#1)}}}}

    \newenvironment{myquote2}{}{}

     \newcommand {\revised}[1]{{#1}}

\usepackage{array}
\newcolumntype{L}[1]{>{\raggedright\let\newline\\\arraybackslash\hspace{0pt}}m{#1}}

\begin{document}

\title{Creativity in the BioFoundry: Supporting scientific creativity in the age of automation}

\author{Mingyan Claire Tian}
\email{tian26@illinois.edu}
\orcid{0009-0004-7395-1429}
\affiliation{%
  \institution{University of Illinois Urbana-Champaign}
  \city{Urbana}
  \state{Illinois}
  \country{USA}
}

\author{Sarah Sterman}
\orcid{0000-0002-9282-559X}
\email{ssterman@illinois.edu}
\affiliation{%
  \institution{University of Illinois Urbana-Champaign}
  \city{Urbana}
  \state{Illinois}
  \country{USA}
}

\renewcommand{\shortauthors}{Tian and Sterman}


\begin{teaserfigure}
    \centering
  \includegraphics[width=0.8\textwidth]{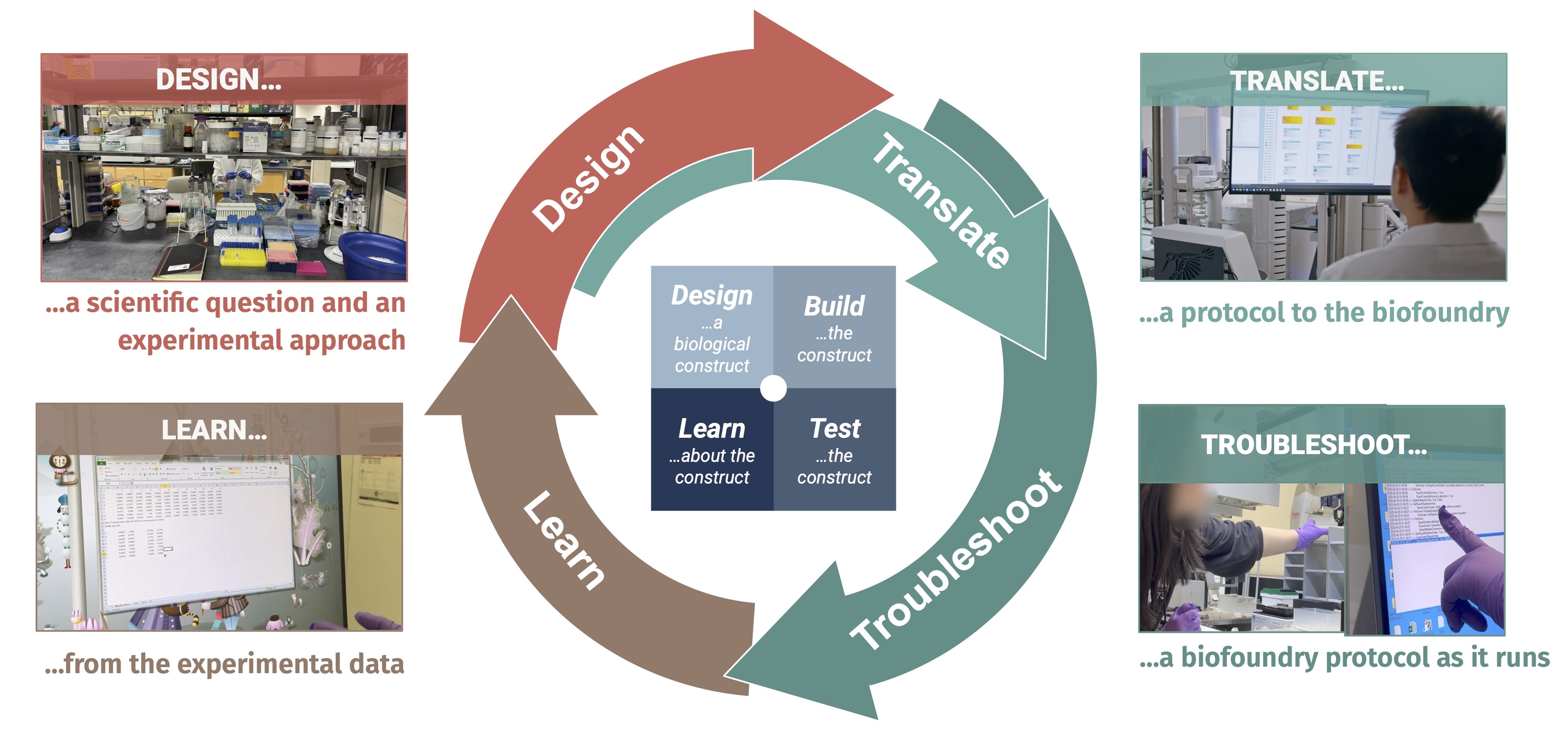}
  \caption{\revised{Scientific Creativity in Biofoundries: Creative Actions Cycle. Biofoundries are automated laboratories for running biology experiments. They change where and how creativity occurs during experimental science, adding new considerations to the Design-Build-Test-Learn experimental cycle (center, inner quadrant). Creativity in biofoundries include experimental design, translation into machine-executable workflows, troubleshooting failures, and learning from results, reflecting how automation redistributes creative labor. We describe the new forms of creativity involved in working with biofoundries using the Design-Translate-Troubleshoot-Learn (outer cycle) as an additional layer to the DBTL cycle.}}
  \Description{Annotated image of the design-build-test-learn cycle in the biofoundry, with sites where creativity occurs. This includes designing the scientific question and experimental architecture, translating them to automation, troubleshooting, and learning from the outputs and process to update the scientific question and experiment.}
  \label{fig:teaser}
\end{teaserfigure}


\begin{abstract}

Biofoundries automate biological experimentation at unprecedented scale, promising speed, reproducibility, and access. Yet automation also reshapes how scientists experience experimentation and creativity.
Through in-depth interviews with nine scientists and experts across academia and industry (including biofoundry developers, automation engineers, and end-users), we examine how scientific creativity is enacted under automation. Biofoundries displace sensory cues, redistribute responsibility between humans and machines, and transform troubleshooting from an embodied, local practice into a predictive, social, and interpretive one. Rather than framing biofoundries as automation factories, we argue that they should be understood as Creativity Support Tools, whose design directly shapes how researchers notice breakdowns, exercise judgment, learn from failure, and progress through success. 
By connecting biofoundry practice with prior HCI work on automation, debugging, and distributed creativity, this paper demonstrates biofoundries as a distinctive and timely site for creativity research in science.

\end{abstract}

\maketitle

\section{Introduction}

Automation tools are increasingly central to contemporary scientific practice. They promise speed, reproducibility and scale \revised{by offloading} tedious or repetitive labor \revised{to machines and standardizing experimental procedures (e.g., biofoundry pipelines \cite{ChaoBiofoundry2017}) and cloud laboratories \cite{ArnoldCloudLabs2022}}. However, creativity research has long shown that creative work does not reside solely in moments of ideation or insight \revised{\cite{Sawyer_Henriksen_2024}}, nor can it be easily separated from the material \revised{\cite{cognition}}, social \revised{\cite{Sawyer_2009}}, and interpretive practices \revised{\cite{Schon_1983}} through which work actually gets done. From this perspective, automation does not simply address efficiency, but also reconfigures creativity, \revised{shifting efforts from manual execution to oversight and coordination \cite{BainbridgeIroniesAutomation1983}}, \revised{transforming embodied engagement with materials into interaction through representations \cite{Klemmer_2006}, and reshaping} how practitioners think about their contributions \revised{\cite{Hammang_2023}}.

Scientific work provides a particularly rich context for examining these shifts. Laboratory practice studies describe science as a creative, iterative activity, shaped by breakdowns, repair \revised{(where unexpected results prompt cycles of repair and reinterpretation \cite{Pickering_1995})} and ongoing sense-making rather than linear execution \revised{of predefined plans \cite{Latour_Woolgar_1979}}. As automation becomes more deeply embedded in scientific infrastructures, these practices do not disappear; rather, the conditions under which they occur change. 

\textbf{Biofoundries} 
offer timely lens on these transformations. Commonly framed as infrastructure or platforms for faster, more reliable experimentation at scale \cite{PetzoldBenchtoBIofactory2025}, biofoundries require researchers to design experiments, interpret results, and make sense of failures with reduced sensory access to materials and physical processes. These tensions raise questions about where creative work happens in biofoundry practice, how it is supported or constrained, and how scientists adapt their ways of working in response. 
Answering these questions can reveal opportunities for software tools to support biology researchers, learners, and educators working with biofoundries or other lab automation tools.

In this paper, we approach biofoundries through the lens of \textbf{Creativity Support Tools (CSTs)} \revised{\cite{schneiderman_2009}}. Rather than treating biofoundries solely as technical systems for executing experiments, this perspective foregrounds how tools shape exploration \revised{\cite{Resnick_Robinson_2018}}, learning, interpretation \revised{\cite{Schon_1983}}, and collaboration \revised{\cite{Klemmer_2006}}. It invites attention to the everyday practices through which creativity is enacted, including troubleshooting, coordination with others, and the gradual refinement of experimental ideas (Fig. \ref{fig:teaser}). Our framing does not assume that biofoundries are successful creativity support tools or not. It provides a way to examine what kinds of creativity they currently enable, and where practitioners experience friction or loss.

Drawing on interviews with nine biofoundry practitioners across academia and industry, we make three contributions. First, we characterize how practitioners experience biofoundries as creativity support tools beyond efficiency and throughput. Second, we identify forms of creativity in biofoundry work, and how these differ from creativity in bench-based experimentation, particularly in how creative value is placed on design, anticipation, and interpretation. Third, we examine how biofoundries alter the material conditions of creativity, redistributing responsibility and agency across people and systems. 

In the discussion, we connect these observations to creativity theory and prior studies of automation in domains like digital fabrication and computational design, where researchers have similarly examined how programmable tools reshape \revised{material engagement \cite{Beuchley_2010}, feedback \cite{Klemmer_2006, Landwehr_2022}, and collaboration \cite{Hirsch_2023}}. 
We highlight both familiar patterns and domain-specific challenges, such as the invisibility of biological materials, delayed experimental feedback, and the high cost of failure. Together, this paper suggests that viewing biofoundries as creativity support tools offers a productive way to understand how creative scientific practice is changing, and to inform the design of future systems to better support experimentation, learning, and discovery.

\section{What is a Biofoundry?}

\begin{figure*}
  \includegraphics[width=0.9\textwidth]{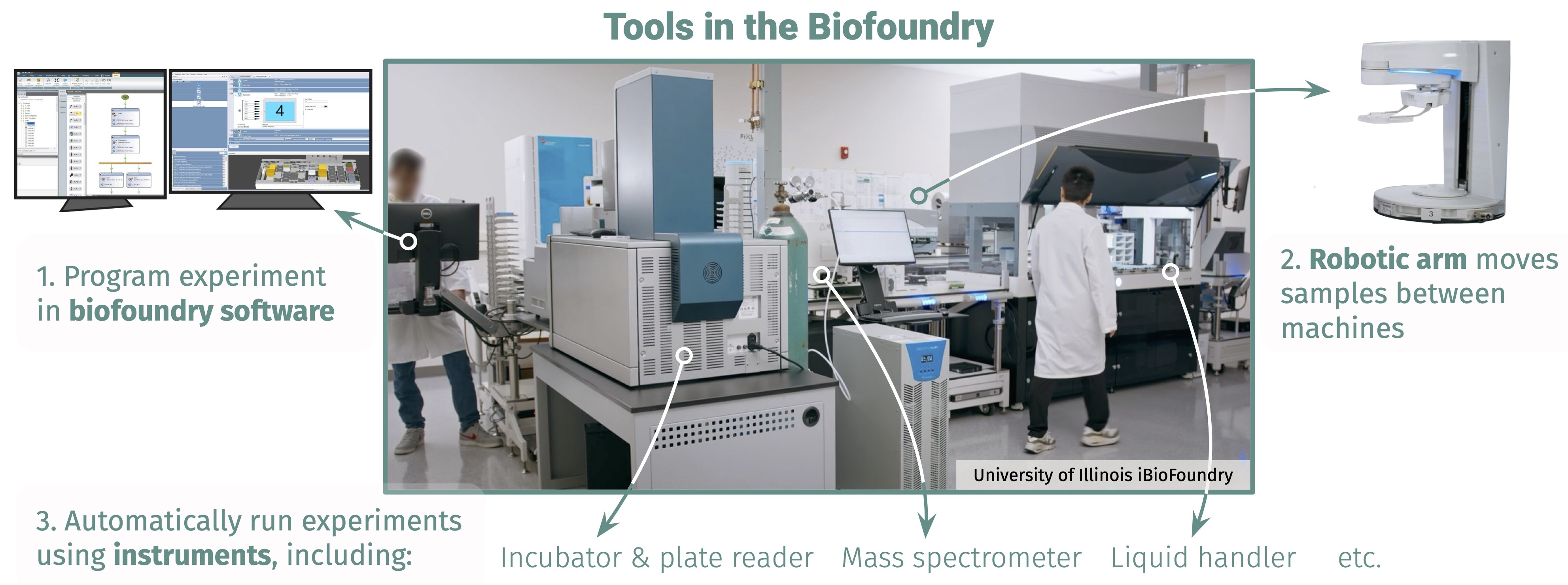}
  \caption{To run an experiment in a biofoundry, an experimental protocol is programmed with software (1), then the physical experiment is carried out by a combination of robotic tooling (2) and specialized instruments (3). The results are returned to the users as logs and other digital outputs.}
  \Description{Instrumentation of biofoundries. Compared to the manual bench, biofoundries typically operate by a central scheduling software that controls the robotic arm moving plates between other experimental or analytical instruments.}
  \label{fig:instruments}
\end{figure*}

A \textbf{biofoundry} is a centralized, highly automated laboratory facility designed to execute biological experiments at scale \cite{ChaoBiofoundry2017}. Typical biofoundries integrate robotic liquid handlers, automated incubators and plate readers, software for workflow specification and scheduling, and centralized data capture and logging (Fig. \ref{fig:instruments}).

Synthetic biology experiments are commonly organized around the \textbf{Design-Build-Test-Learn (DBTL)} cycle \cite{Kim_2025}. Once a question or hypothesis is defined, the DBTL cycle structures the engineering and evaluation of the biological system. In the Design phase, scientists use computational tools to design biological constructs to achieve a particular goal (e.g. producing insulin). In the Build phase, they create the constructs in the physical world (e.g. synthesizing the DNA sequence). In the Test phase, the constructs are tested in the real world (e.g. introduced into living cells). In the Learn phase, the results collected from testing are analyzed to inform subsequent designs (e.g. evaluating expected vs observed insulin production) \cite{Hillson_2019}. In a biofoundry, scientists create the protocols to input into the automated instruments in the Design phase; Build and Test phases are executed on the instruments, often asynchronously and remotely; and results are returned from the biofoundry instruments to the scientists for the Learn phase.
Analysis from the Learn phase may then be used to drive another iteration of the cycle, ideally with modified and improved results. (Fig. \ref{fig:teaser}, inner quadrant)

Over the past decade, biofoundries have moved from speculative visions to operational research infrastructures for automated biology \cite{Hillson_2019}. 
Today, only about 40 academic, and industrial biofoundries exist globally\footnote{https://www.biofoundries.org/} (compared to, for example, thousands of digital fabrication labs in a single network\footnote{https://fabfoundation.org/hq/}). 

Prior literature frames biofoundries primarily as factories for scale, pipelines that reduce human variability, or platforms for democratization, but say little about how practitioners \textit{experience} automation during breakdowns, which are precisely when creativity and judgment are most needed. \revised{For example, \cite{Hillson_2019} explicitly position biofoundries as ``cloud laboratories'' for biological engineering, emphasizing standardized workflows and automation as mechanisms for scalability and reproducibility. \cite{RoschAutobiotech2024,DudleyBiofoundryPlant2021,ArnoldCloudLabs2022} emphasizes their role as infrastructures for distributed experimentation and access.}


Our work builds on, but departs from, these framings by focusing on troubleshooting as a lived, creative practice. For HCI, biofoundries matter not simply as efficient infrastructure, but as \textbf{emerging creativity support systems}. This places biofoundries at a valuable moment of transition, as their workflows, interfaces, and divisions of labor are still being actively negotiated, making visible how automation reshapes creative work across people, software, and machines \cite{cognition,Suchman_1987}. 
Studying biofoundries now offers a rare opportunity to examine creativity \textit{as it is being reconfigured}, before these practices solidify.
\section{Related Work}

\subsection{Creativity in Scientific Practice: Iteration, Failure and Repair}





\begin{table*}[t]
    \centering
    \begin{tabular}{|L{0.11\linewidth}|L{0.13\linewidth}|L{0.33\linewidth}|L{0.09\linewidth}|L{0.2\linewidth}|}
   \hline
    \rowcolor{lightgray} \textbf{Participant} & \textbf{Biofoundry Role} & \textbf{Experience} & \textbf{Industry/ Academia} & \textbf{Domain} \\
    \hline
        P1 & End-user & Built 1 workflow & Academia & Microbial engineering \\ \hline
        P2 & Founder & Developed a biofoundry, led 2 projects & Academia & Crop science \\ \hline
        P3 & Founder & CEO of a cloud biofoundry & Industry & Quality control \\ \hline
        P4 & Engineer & Led 9 projects, mentored >10 & Academia & Microbial engineering \\ \hline
        P5 & Engineer & Senior lab technician & Industry & Bio-production \\ \hline
        P6 & Founder & Designed a biofoundry, led 3 projects & Academia & Protein engineering \\ \hline
        P7 & End-user & Graduate student in training & Academia & Directed evolution \\ \hline
        P8 & End-user & Conducted 3 projects & Academia & Metabolic engineering \\ \hline
        P9 & End-user & Conducted 6 projects & Academia & Biosynthesis \\ \hline
    \end{tabular}
    \caption{Interview participants covered a range of expertise and roles in biofoundries.}
    \label{tab:participants}
\end{table*}

Creativity research increasingly characterizes science as a creative practice grounded in everyday work rather than rare moments of discovery \cite{Sawyer_Henriksen_2024,Glaveanu_2014}. Scientific creativity emerges through cycles of hypothesis formation, experimental construction, and interpretation of ambiguous results \cite{Nersessian_2006}, grounded in material engagement and informal reasoning rather than formal planning alone \cite{Latour_Woolgar_1979,Schon_1983, Pickering_1995}. 
In this perspective, creativity is distributed across people, instruments, representations, and materials, aligning with HCI accounts of creativity as a situated and socio-technical practice \cite{Sawyer_2009,Suchman_1987,Frich_2019}.

A central mechanism of creativity in scientific practice is iteration through breakdown and repair. 
\revised{Ethnographic studies show that experiments rarely proceed as planned, but advance through cycles of failure, adjustment, and reinterpretation \cite{Pickering_1995, Latour_Woolgar_1979}.} Scientific creativity is shaped by what tools make visible or invisible, \revised{what actions are easy or costly \cite{FegerRoleHCI2019}}, and how representations support sensemaking over time \cite{Bodker_2015}. HCI researchers also emphasized that creativity in technical domains is inseparable from tools and infrastructures. 
\revised{In computational domains, debugging and error recovery are framed as key sites of creative production rather than mechanical correction, as they have been shown to involve iterative hypothesis generation and testing \cite{Kery_myers_2017} where practitioners interpret system behavior and refine their understanding of underlying processes \cite{Ko_2015}.}
From this perspective, creativity is not only a cognitive act but also a distributed accomplishment, spanning people, artifacts, and institutional arrangements \cite{Suchman_1987}. However, much of this work has focused on computational or observational sciences, leaving experimental life sciences, and especially automated biology, less examined as sites of creative practice. As advances in biotechnology and high-throughput experimentation promise to increase scientific creativity by enabling larger experimental spaces and more systematic comparison \cite{ChaoBiofoundry2017,PetzoldBenchtoBIofactory2025}, it is our core concern to understand this shift in scaling and speed and how it reshapes when, where, and how creativity occurs.

\revised{The growing intersection of the life sciences and HCI uses diverse lenses for understanding participation in science. Prior work in BioHCI has explored the democratization of biotechnology through the DIYbio movement \cite{Kuznetsov_2015}, framing open-source biology tools as platforms that catalyze hybrid knowledge production by lowering the bar for public participation. Hamidi et al. explore ``transdisciplinary fluency'' in bioart, where interacting with DNA becomes a site for symbolic and ethical exploration \cite{Hamidi_2021}. Chen et al. take a more-than-human lens to address the ``labor provenance'' of microbes in synthetic biology \cite{Chen_2025}. By studying biofoundries through the lens of its role in creativity, we explore the relationship between HCI and synthetic biology in a high-cost, centralized, and highly abstracted environment where professional scientists' agency and creativity are being reconfigured.} 



\subsection{Automation and Creativity Support} 
Creativity Support Tools (CSTs) aim to amplify creative practice by supporting exploration, iteration, and reflection rather than replacing human judgment \cite{shneiderman,Frich_2019}. Across creative domains, automation is known to relocate rather than eliminate creative labor. Studies of automation highlight how human effort shifts from direct manipulation toward monitoring, interpretation, and exception handling \cite{BainbridgeIroniesAutomation1983}. 

In creative work, this often elevates troubleshooting as a key site of insight generation, where breakdowns expose assumptions and prompt new understanding \cite{Suchman_1987,Schon_1983}. Studies across HCI document how automation introduces new forms of breakdown that demand creative sensemaking \cite{Norman_2013, Amershi_2019}. In programming and AI-assisted tools, users must interpret logs, traces, and partial feedback to diagnose failures, often engaging in collaborative debugging, showing that automated systems introduce new cognitive demands related to trust, interpretability, and control \cite{Ko_2015,Amershi_2019,HeerAgencyAutomation2019, Beuchley_2010}. Rather than removing human judgment, automation requires users to reason about system behavior, assess reliability, and decide when to intervene.

Digital fabrication offers especially relevant parallels. Programmable workflows such as CNC machining and 3D printing decouple designers from direct material engagement \revised{\cite{Vallgrarda_2007}}, altering sensory feedback and iteration cycles \cite{Klemmer_2006}. Prior work shows how makers compensate through simulation \cite{Hirsch_2023}, material exploration \cite{Twigg_2023}, and partial test pieces \cite{Landwehr_2022,Subbaraman_2025} to restore creative control. These studies highlight trade-offs between abstraction and flexibility, speed and cost, that shape creative exploration under automation. These dynamics suggest that creativity under automation depends not only on algorithmic capability, but also on how systems support interpretive engagement and learning. 
\revised{By placing biofoundries in conversation with these domains, we highlight both familiar patterns of material probing and domain-specific challenges, such as the high cost of biological failure and the unique opacity of living materials.} 




Our study connects these threads by examining how \textbf{automation reshapes the creative act of experimenting in science}, extending C\&C conversations into bioengineering.

\section{Methods}

\subsection{Data Collection}
We conducted semi-structured interviews with nine participants (N=9) involved in the design, operation, and use of biofoundries across academia and industry. Participants included 3 biofoundry founders, 2 automation engineers, and 4 end-users (Table \ref{tab:participants}). These roles reflect different relationships with biofoundries, from building and maintaining workflows to encountering them as part of their everyday research, allowing us to examine how creativity and judgment are distributed across biofoundry practice.

Interviews were conducted over Zoom and lasted approximately 60 minutes. They focused on everyday interactions with biofoundries, with particular attention to moments of uncertainty, breakdown, and decision-making. Prompts explored interaction challenges, mental models of automated workflows, learning practices, and how participants described creativity in their work, encouraging participants to surface concrete episodes they found illustrative rather than abstract evaluations of automation.

\revised{While our participants span distinct roles, we did not treat these as separate analytic groups. Instead, we used role differences as a source of analytic contrast, attending to where perspectives aligned and where tensions emerged. In particular, differences in how participants defined ``creativity'' often reflected their proximity to execution (e.g., engineers emphasizing reliability, end-users emphasizing control, founders emphasizing scale). We return to these tensions in Findings (Sec. \ref{sec:findings}) and Discussion (Sec. \ref{sec:discussion}).}

\subsection{Analysis}

We performed a reflexive thematic analysis \cite{braunReflectingReflexiveThematic2019,braunUsingThematicAnalysis2006}. Interviews were transcribed with minor edits for grammar. The authors began with open coding that focused on concrete practices (e.g., low level descriptive codes like \textit{watching robot runs}, \textit{running ``safe'' experiments}, and \textit{consulting core staff}). Codes were iteratively clustered into higher-level themes through memoing. We reapplied the higher-level memos to the corpus and refined the codes and memos. We constructed themes that characterized how creativity, troubleshooting, and interpretation are enacted in biofoundry work. This practice-centered approach aligns with our goal of understanding creativity as something that unfolds through situated interaction with automated systems, materials, and people.

\section{Findings}
\label{sec:findings}



Across interviews, participants described biofoundries 
as environments that reorganize where creativity happens, who participates in it, and what kinds of scientific contributions are valued. Rather than a simple story of automating away tedium, biofoundry work is deeply entangled with material constraints, social coordination, and interpretive labor. In this section, we answer our two main research questions: \textit{Where do scientists locate creativity in their work with biofoundries?} and \textit{How does the biofoundry alter the material conditions of creativity?}



\begin{figure*}[h!]
  \includegraphics[width=0.9\textwidth]{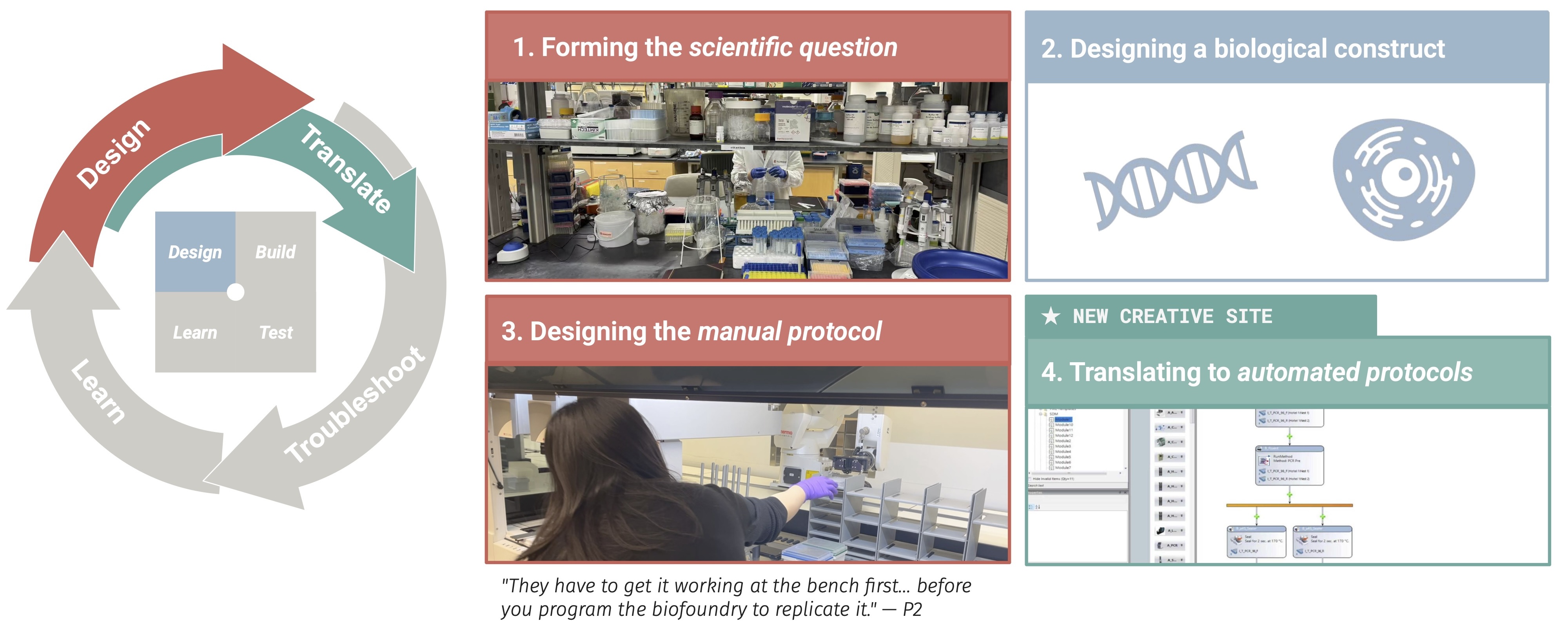}
  \caption{\revised{In the \textit{Design} phase of the DBTL cycle (Sec. \ref{sec:design}), scientists form the scientific questions and design a manual protocol. When a biofoundry is used, there is a new type of creative work distinctive to biofoundries added: translating the protocol into robot-executable logic. The automated protocol is itself a creative artifact: it encodes decisions about contingencies, timing and material constraints that must be anticipated in full before execution begins. 
  }}
  \Description{The Design phase of the DBTL cycle is highlighted to bring out four sites where creativity occurs in this phase. The first site is framing the scientific question, accompanied by an image of the lab bench. The second site is designing the biological construct, accompanied by the icon of a cell and DNA. The third site is designing the manual protocol, accompanied by an image of a researcher manually retrieving plates in the lab. The fourth site is translating to automated protocols, accompanied by a screenshot of an example protocol a researcher developed in biofoundry software. This fourth site is marked as a new creative site that only exists in biofoundries.}
  \label{fig:design}
\end{figure*}

\subsection{Where do scientists locate creativity in their work with biofoundries?}

The standard \textit{Design, Build, Test, Learn} (DBTL) approach to biology experiments divides the scientific work into \textit{designing}, i.e., coming up with the experimental question, \textit{building} a protocol or test set up, \textit{testing}, i.e., executing and troubleshooting the protocol, and \textit{learning}, i.e., analyzing and interpreting the data to update one's understanding of the question and experiment to feed back into the next cycle.  While the design and learn phases may appear to be the most clearly creative, in that these are where the questions are being asked and answered, creativity is essential through all phases of the cycle. Participants consistently described creativity in biofoundries as an everyday, situated practice embedded across multiple stages of experimental work, rather than as isolated moments of ideation. However, compared to benchwork, automation shifts where effort and ingenuity are required, foregrounding design, prediction, and interpretation over hands-on manipulation. In addition, different scientific roles shift where they locate the key moments of creativity.

\begin{figure*}[t]
  \includegraphics[width=0.9\textwidth]{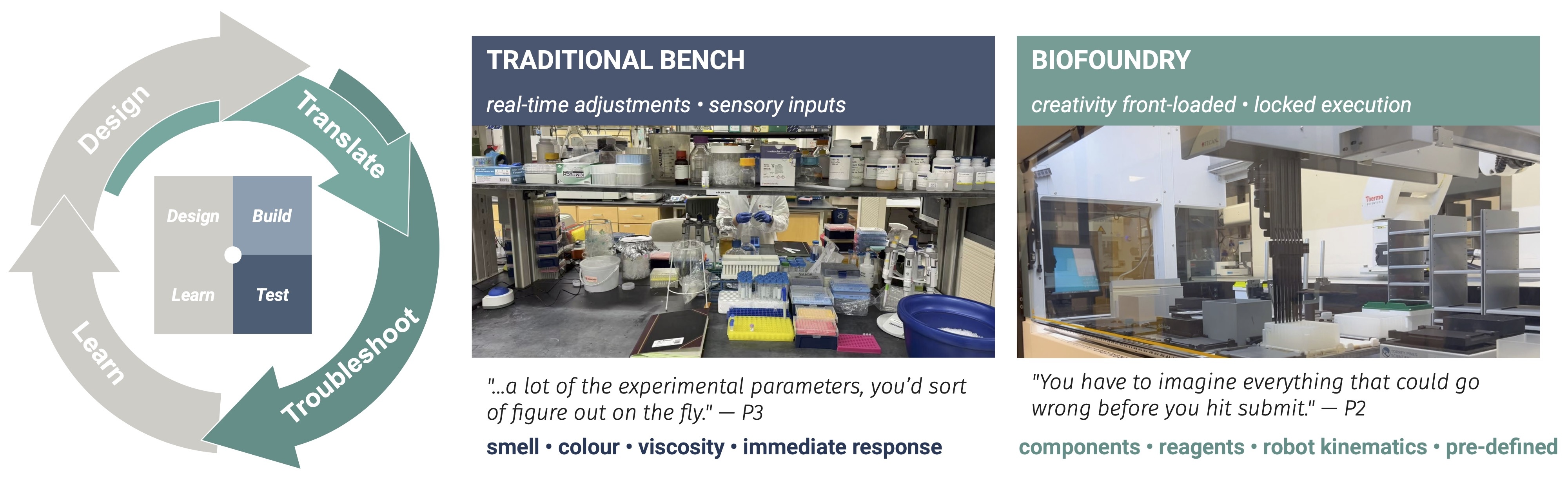}
  \caption{\revised{In traditional bench science, creativity is stretched out continuously across the \textit{Build} and \textit{Test} phases (Sec. \ref{sec:buildtest}) as real-time sensory adjustment. Under biofoundry automation, the execution becomes a largely uninterruptible window as intervention becomes costly once the run begins. Creative effort is displaced upstream into intensive anticipatory design.}}
  \Description{The Build and Test phase of the DBTL cycle is highlighted. We draw a comparison between how creativity occurs at the traditional bench versus in a biofoundry during these two phases. At the bench, an image of a cluttered workstation and notepads show that this mode allows for real-time adjustments and sensory inputs. For the biofoundry, a robot liquid handling executing pipetting commands onto a plate illustrates that scientists must imagine everything that could go wrong before they start execution.}
  \label{fig:buildtest}
\end{figure*}

\subsubsection{Design: Creativity is in Designing the Question and the Experimental Architecture}\label{sec:design}

Some biofoundry scientists drew a sharp distinction between the \textit{design} phase of the DBTL cycle as where the creative intellectual contribution resided, and the \textit{build} and \textit{test} phases as simply execution. This perspective was prominent among participants from industry biofoundries; P3, now the head of an industry biofoundry, drew such a distinction:

\begin{myquote}
    ``Designing... what is going to happen when, and then analyzing the data and metadata that comes out of it, to me, that's the scientific part. The... execution part is actually going through those steps in the laboratory.'' \participant{P3}
\end{myquote}

For industry participants like P3, biofoundries clarify what ``counts'' as science, and refocus scientists on what matters. P3 contrasted this with his experience in traditional wet labs, where a long apprenticeship centered on developing manual skill and tacit know-how can leave scientists equating their scientific value with what they can physically do at the bench.
From this perspective, creativity and scientific value is meant to reside in \textit{designing the experimental question}, and biofoundries enable this shift. 

However, a distinction between intellectual design and manual execution is not always so clean. Among other participants, designing the protocol needed to run the experiment in the biofoundry was also considered a creative part of the design, requiring significant testing, iteration, and manual work. 
Many participants described first designing the protocol at the bench with manual methods, before translating to the biofoundry:

\begin{myquote}
    ``
    We decided firstly on how we are going to engineer the proteins, and then once we have a bench workflow set up, then we created a robotic workflow for all the bench steps.''
    \participant{P6}
\end{myquote}

\begin{myquote}
    ``[The researchers] have to get it working first... 
    before you then program the biofoundry to replicate what that student is doing.''
    \participant{P2}
\end{myquote}

After the manual work of designing the protocol, participants also described the act of making an experiment work on a robot, reliably, at scale, as an intellectual contribution in its own right. Several researchers characterized the automated workflow itself as the ``fruit'' of their creative labor, emphasizing it as an artifact embodying decisions about timing, volumes, contingencies, and constraints (P4, P6).

\begin{myquote}
    ``The experimental design is also my creation... 
    Beside the hypothesis.
    Making a path to test that and actually get it to work...
    is definitely a huge portion of contribution.''
    \participant{P6}
\end{myquote}

P5, an automation engineer at an industry Biofoundry, distinguished between scientists in academia who see themselves as ``artists'' whose experiments are precious and creative to them, and scientists in industry (primarily working with quality control) who adopt a ``factory approach,'' prioritizing standardized outcomes. Among the ``artists'', fulfillment and pride was tied to the creativity in the experimental design and execution. Such participants described a desire to feel connected to and in control of the process of discovery, and for some, biofoundries threatened this connection. P7, an academic researcher, felt that ``the research... loses a little bit of the kind of pride that I have when I do an experiment'' when the biofoundry does it all.  
P7 contrasted this unease with using other machine tools, like bioreactors for tracking cell growth, that did not take away the emotional connection and was still ``fulfilling.''


Yet on the other side of the spectrum, industry participants said some biofoundry users would prefer to offload that part of the design to an automation expert:

\begin{myquote}
    ``These are usually scientists at big pharma companies. They don't just want the race car, they also want the person to drive it. So we've actually found a lot more commercial traction by building an internal solutions team that will work with customers to transfer [their methods] or to develop [new ones].
    ''
\participant{P3}
\end{myquote}

\begin{myquote}
   ``It sounds very patronizing, but 
   they want someone else sitting behind the wheel... 
   if that's what comes out of the box, I don't really care what's in the box.'' 
    \participant{P5}
\end{myquote}

The creativity in the scientific endeavor was a source of pride and fulfillment for many participants, and shaped their identities as scientists. But where that creativity lay depended on their goals and roles. 
\revised{Participants' roles did not simply influence how they perceived creativity, but also structured where creativity could occur within the biofoundry system. End-users located creativity in control, intuition, and connection to experimental outcomes, often experiencing automation as distancing or constraining. Engineers and biofoundry developers framed creativity in terms of system design, reliability, and workflow construction. Founders emphasized abstraction and the separation of ``science'' from execution. Rather than contradictions, these perspectives reflect a system in which creativity is distributed across roles, and what counts as creative work depends on one's position within the infrastructure.}





In summary, transitions to biofoundries emphasize pre-planning experiments and focusing on the question as the site of creativity, but designing a question is still entangled with the details of designing a protocol, both through preparatory bench work and the material constraints of the automated system (Fig. \ref{fig:design}). This leads to a new site of creativity, in the design of the automated protocols themselves. The goals and role of the scientist, whether seeking to create new knowledge about how things work or seeking to reach a particular outcome regardless, shape their site of creativity.



\subsubsection{Build + Test: From Fiddling to Front-loaded Creativity}\label{sec:buildtest}

A central contrast participants drew between benchwork and biofoundry work concerned when creative decisions are made. Traditional bench science was repeatedly described as an ``organic process,'' where parameters are adjusted on the fly in response to immediate sensory feedback (Fig. \ref{fig:buildtest}). P3 vividly recalled this
mode of working:

\begin{myquote}
    ``You'd get into the lab in the morning, and you'd have a vague idea of what experiments you wanted to do... 
    you'd sort of figure out on the fly.'' \participant{P3}
\end{myquote}


P6 described how experience and past knowledge meant very little planning was needed. When doing PCR manually, he does not consciously think about the steps, instead ``[I] copy from all the other times I've done it before.''
In contrast, the biofoundry required more planning, separating the experimental design and the experimental execution. 

\begin{myquote}
    ``[Moving the protocol to the biofoundry] will have to use two equipments...So instead of doing that, what I did was I [recalculated and] diluted my primers, and set up everything on one machine...that's something you need to think, plan. For most of the experiment, you have to work through each step, which you don't really think about a lot before.''\participant{P6}
\end{myquote}

\begin{figure*}
  \includegraphics[width=0.9\textwidth]{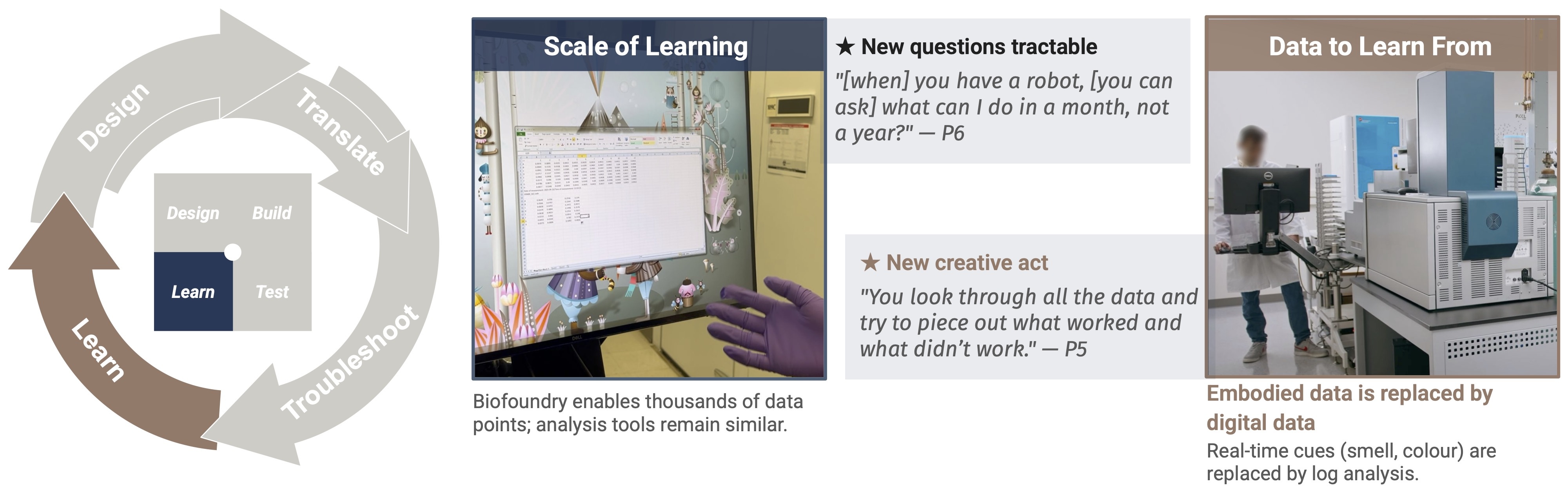}
  \caption{\revised{The \textit{Learn} phase (Sec. \ref{sec:learn}) bifurcates into two structurally distinct streams under automation. Learning about the scientific question changes mainly in scale, as the same analytical approaches now operate over orders-of-magnitude more data, enabling new categories of inquiry. Learning about the experiment itself is more profoundly disrupted, as embodied, real-time cues are replaced by retrospective interpretation of metadata logs, demanding a new form of interpretive reasoning that combines programmatic analysis with tacit bench expertise.}}
  \Description{The Learn phase of the DBTL cycle is highlighted. An image of an output CSV file is put in parallel with a scientist working with biofoundry software. This highlights that both learning in terms of science and learning about the experiment is changed under biofoundry automation.}
  \label{fig:learn}
\end{figure*}

Working with a biofoundry means you cannot do real-time tinkering. Once a biofoundry run begins, scientists cannot ``stick [their] hand in'' mid-execution to ``turn that knob a little bit more'' (P5). As a result, creativity is pushed out of the \textit{Test} phase, and upstream to the design phase. Participants described having to imagine and encode every possible contingency in advance, a process several framed as both intellectually demanding and creatively generative:

\begin{myquote}
    ``You have to imagine everything that could go wrong before you hit submit. With the biofoundry you spend more time upfront imagining what might go wrong.
    \participant{P2}
\end{myquote}

Researchers moving from bench to biofoundry need to translate informal, experience-based protocols into explicit, machine-readable logic. While a person can adapt on the fly if there is a mistake or new information, a robot cannot. This means scientists must emphasize the ``logic'' of automation.

\begin{myquote}
    ``In theory, that protocol that we write in our laboratory notebooks or papers ought to be the equivalent of pseudocode. 
    In practice, there's nothing to make sure that actually happens. So what tends to happen is that the person who's doing it writes down everything they need to remember to do. And that's different.''
\participant{P2}
\end{myquote}

Participants emphasized that translation to the biofoundry was not merely mechanical. P6 described protocols as being ``lost in translation,'' requiring creative adaptation to robotic constraints such as minimum volumes, reagent handling limitations, or movement efficiency. These adaptations sometimes reshaped scientific decision-making itself, with researchers prioritizing, for example, ``less movement on the robot'' over experimental configurations they might have chosen at the bench.

Moving the protocol design upstream and away from manual intervention also does not remove it from material considerations.  In some ways, the material realities become more present and complex. On the bench, a scientist might rely on their past knowledge, available tools, and respond in real time to a material challenge. In the biofoundry, those challenges must be considered ahead of time.  In one of many examples, P2 discussed the effect of the type of plastic in a pipette on the growth of cells:

\begin{myquote}
    ``You measure [a solution] using these robotic pipetters with these little disposable plastic tips on them. And those are not standardized... they'll use slightly slightly different plasticizers. 
    And so, you can change to a different brand of tips, and suddenly your cells don't grow, or they grow too well.''
    \participant{P2}
\end{myquote}

In summary, in biofoundries, researchers cannot adapt the experiment while it's running. To handle that constraint, researchers attempt to do more upfront planning of the experiment. 


\subsubsection{Learn: Updating the Scientific Question and Experiments}\label{sec:learn}

The final phase of the DBTL cycle, \textit{learn}, addresses both learning about the scientific question, and learning about the experimental design (Fig. \ref{fig:learn}). 

Learning about the \textit{scientific question} seemed to change the least for our participants. This may change in the future, as biofoundry designers seek to integrate more AI-enabled analytical tools, but with the tools available in our participants' biofoundry experiences, analyzing data remained a similar process. It primarily happens at the end of a test cycle, using suites of bioinformatics software:

\begin{myquote}
    ``It's not that you can notice it at the bench, this is some data that you see later on...
     ''
     \participant{P9}
\end{myquote}

\begin{figure*}[h!]
  \includegraphics[width=0.9\textwidth]{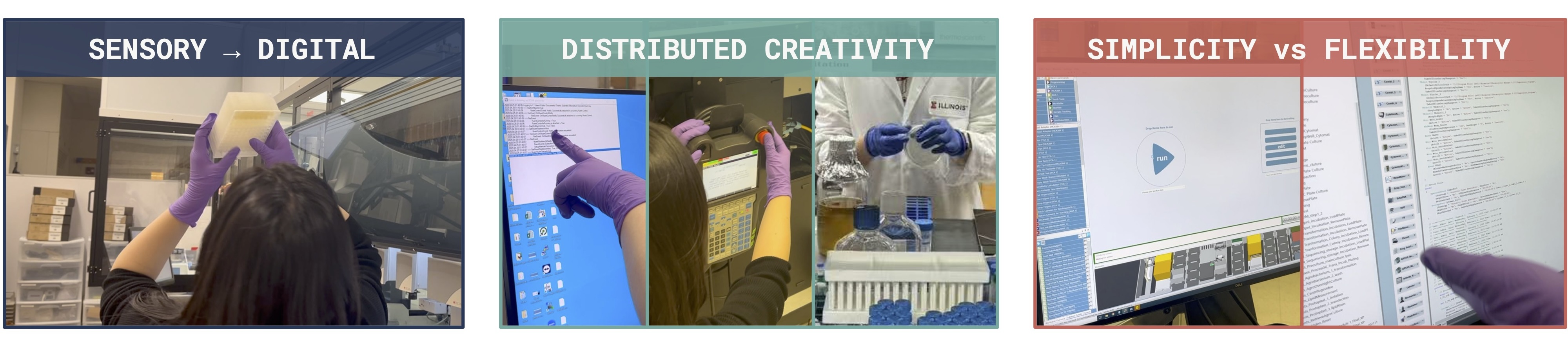}
  \caption{\revised{Biofoundries reshape the material conditions of creativity. Left: sensory information, such as a scientist checking the color of wells, is replaced by digital signals. Middle: creative labor is redistributed across scientists, staff, and machines. Right: interface designs surface tensions between operation simplicity and scientific flexibility.}}
  \Description{The left panel shows the shift from sensory to digital signals with photo of a scientist visually checking the cultures in a plate; the middle panel shows social creativity distributed across roles with core staff helping the scientist troubleshoot through system logs, machine knobs, and bench work; the right panel shows the simplicity-flexibility tradeoff with a picture of a ``big green button'' interface design.}
  \label{fig:condition}
\end{figure*}

The main change to the scientific question analysis appeared in the scope of what participants were able to study. P6 and P4 described moving from hundreds to thousands or tens of thousands of data points, making previously unimaginable questions tractable. As P6 put it:


\begin{myquote}
    ``The guiding principle of your creativity has to be the scientific question...[when] you have a robot, [you can ask] what experiment I can do in a year, now can we do it in one month? And since I can do this experiment in a month, so can I do 10 more in the year?''
    \participant{P6}
\end{myquote}

Not just the scale itself changes, but types of questions that are impossible to answer without scale also become tractable (P4). This change appears both in the initial design of a scientific question, and in the analysis of the data made possible by the biofoundry. But the type of question enabled by the biofoundry remains entwined with the creativity of how to effectively use the automation tools:

\begin{myquote}
     ``Making the experiments on the robots is also going to take a lot of tweaking, a lot of programming, a lot of creativity...the biofoundry part is a different type of creativity. 
     you are finding the \textit{how} [to do] things for a scientific question.''
    \participant{P6}
\end{myquote}

In contrast, learning about the \textit{experiment} changed significantly. Rather than learning through embodied interaction with materials during the build and test phases, participants described learning through post-hoc analysis of large datasets and metadata logs. P5 contrasted these modes:

\begin{myquote}
    ``Instead of fiddling in real time, you run an experiment... 
    you look through all the data and metadata and try to piece out what worked and what didn't work, and what changes you need to make next time.'' \participant{P5}
\end{myquote}

For some participants, this cleaner division encouraged a more systematic and reflective form of scientific creativity, even as it departed from familiar practices.  But such divisions also delay learning, such as when direct contact with the experiment can indicate early a failure in the protocol. P9 recounted how the sense of smell (yeast smelling like ``beer'' versus ``garbage'') used to alert him of failures whenever he opens the incubator, yet stepping away from the experiment in biofoundries means that you might just get the end result:

\begin{myquote}
    ``
    It's harder to backtrack where it failed. Was it the something in the environment? The incubation? The design? Or did someone in the lab knock over something by accident? You don't know.''
    \participant{P9}
\end{myquote}

In summary, the creation of new knowledge in the biofoundries occurs for both the scientific question and the experimental design, but moves when the analysis of experimental design happens to the end of an experiment.

\subsection{How does the biofoundry alter the material conditions of creativity?}

In the prior section, we presented how biofoundries change the location, type, and uses of creativity in biological experimentation.  Here, we turn to the specifics of how biofoundries act to support and constrain creativity by \textit{changing troubleshooting signals from sensory to digital data}, \textit{changing the social dynamics of support}, and \textit{specifying tool abstractions.} (Fig. \ref{fig:condition}).

\begin{table*}[]
    \centering
    \begin{tabular}{|L{0.1\linewidth}|L{0.36\linewidth}|L{0.46\linewidth}|}
    \hline
        \rowcolor{lightgray}
          & \textbf{Sensory Cues (Manual)} & \textbf{Metadata Traces (Biofoundry)} \\ \hline
        \textbf{Time} & \textbf{Immediate.} Problems are noticed and fixed ``on the fly''. & \textbf{Retrospective.} Data may be analyzed days or weeks later. \\ \hline
        \textbf{Type of Data} & \textbf{Tacit/Holistic.} Based on ``gut feelings'', tacit knowledge and observational skills. & \textbf{Granular/Quantitative.} Based on CSV files, OD measurements, and pressure charts. \\ \hline
        \textbf{Social Context} & \textbf{Centered on the scientist.} The researcher is ``behind the wheel'' and intervenes. & \textbf{Distributed.} Responsibility shifts to the machine or the staff ``babysitting'' it. \\ \hline
        \textbf{Trust Factor} & \textbf{Security.} Physical presence provides ``peace of mind''. & \textbf{Paranoia.} Lack of visibility and the inability to make real-time, tactile adjustments based on ``gut feelings'' or visual cues can lead to a sense of ``helplessness'' or ``disconnectedness''. \\ \hline
        \textbf{Skill Development} & \textbf{Observational Skills.} Soft senses are ``soft abstract concepts'' developed through years of manual experience (such as knowing by eye that a culture is a ``hit''). & Digital troubleshooting requires a \textbf{different skill set}: coding, anticipation, and digital debugging. While some prefer the robot because manual pipetting is ``tiring,'' others argue that losing these soft senses might create ``fragile scientists'' who lack the fundamental biological intuition to spot a failure before a week of work is wasted. \\ \hline
    \end{tabular}
    \caption{In our interviews, we identified two main types of data that could inform troubleshooting. First, ``sensory'' data, primarily in the manual condition, and second ``metadata,'' primarily used in automated biofoundries. These two types of data differ in ways that have implications for the way we approach creativity in biofoundries. Scientists also combine these, and bring skills from one to the other.}
    \label{tab:troubleshooting-compare}
    \Description{This table presents the main ways troubleshooting practice is different in manual versus automated situations.}
\end{table*}

\subsubsection{Sensory signals in troubleshooting} Whether at a traditional lab bench or in a biofoundry, the design and execution of an experimental protocol requires frequent troubleshooting.  Protocols rarely work immediately, whether or not they have been used in the past or are being developed. Troubleshooting is not only a necessary enabler of the creativity of biological experimentation, but also a creative prcoess in itself, where scientists navigate underspecified problem spaces and find new solutions.
Within the biofoundry, troubleshooting changes in time and type, as discussed above, but it also relies on different material realities.  One key change is the shift from sensory cues to digital data in noticing and diagnosing problems.

In traditional labs, researchers use ``soft senses'' like smell, color, and texture to detect failures like contamination. 

\begin{myquote}
    ``If you're doing it yourself, you learn... with your eye that this density is looking positive, and then you're likely to have a hit...'' 
    \participant{P7}
\end{myquote}

\begin{myquote}
    ``After you've done this purification many times, you start to learn compactness of the cells...the color... So, if you're looking at your centrifuge tube, ...[you] immediately know.''
    \participant{P9}
\end{myquote}

The absence of immediate sensory feedback (e.g., seeing culture color, feeling viscosity, noticing growth patterns) was repeatedly cited as a challenge in the biofoundry, particularly when it comes to troubleshooting:

\begin{myquote}
    ``One thing about hands-on experiment is that you can see the culture is off, like the color's weird... 
    Not with the biofoundry... at least not without rapid intermittent human intervention.'' \participant{P2}
\end{myquote}

\begin{myquote}
   ``
   If you're present, you can immediately see it fail, you know what needs to be fixed. Whereas if you just come back and you see that it has failed, but you don't know where, then it's a lot more tedious of troubleshooting.''
\participant{P7}
\end{myquote}

Without the direct access to sensory cues, scientists shift to relying on the digital data created by the biofoundry: metadata, logs, and specifications (e.g., ``pressure traces''(P3) or ``CSV data dumps'' (P8)). In practice, troubleshooting in a biofoundry often involves ``root cause analysis'' of system logs to locate and understand failures (P5). P3 emphasizes that logs can be more detailed in terms of quantity and specificity of data.
But simply having access to more data for troubleshooting isn't necessarily helpful.  P3 goes on to note that biofoundry users frequently don't use the logs to troubleshoot:

\begin{myquote}
    ``Maybe the plate comes back and the fluorescence look like garbage... 
    you can look at, say, the pressure traces of each pipetter... 
    this pipetter clearly went bad, because the pressure just drops precipitously...the end-users they have all the data. They could do it all themselves. But they don't. They just throw up their hands and complain to their customer account manager  
    ...and we figure it out, and then do it again, and it works.'' (P3)
\end{myquote}

Part of this challenge is users' unfamiliarity of with the types of data provided for troubleshooting, as the programming and analytic skills to make sense of machine log data are complex and take time to develop.  Some participants argue that the future of biofoundries is in biologists developing these skills: 

\begin{myquote}
    ``The goal of building a car is not to make people feel like they're still riding a horse... and I do think it requires a very different mentality and cultural and change...''
    \participant{P3}
\end{myquote}

On the other hand, the tacit skills are still essential to successful debugging in biofoundry work. Not all failures will show up in the logs like a worn out pipette's drop in pressure.  
The scientists must be able to combine both the programmatic signals with their experience on the bench and knowledge of how certain problems manifest in the biological system. For example, P6 describes a situation where they need to heat bacteria to a particular temperature: 
\begin{myquote}
    ``Generally, when I do it [at the bench], I have water bath. I just put the plate in there. But now, on the robot, we don't have water bath, we have those metallic plates...
    when you are doing it on the robot, there's a possibility that, because the robot is just [drops] the plate on the heater, so there is a gap. And then another issue is, so these plates are made by some other company... There is not a 100\% snug, tight fitting in there [between the plates and the heater]. 
    So, since I have done experiments [manually], if you do it the first time on the robot, it fails, one of the things I will think of is, hey, did it really get heat at 42 [degrees] or not?'' 
\participant{P6}
\end{myquote}

Nominally, the logs would indicate a correct temperature was set; it requires knowledge of the material situation to notice and debug the incorrect heating of the sample. 

In summary, troubleshooting in the biofoundry removes direct access to sensory cues, requiring scientists to learn to interpret post-hoc log data and perform programmatic debugging.  At the same time, significant expertise is encoded in the manual experience and sensory data gained at the bench that is necessary for successful debugging.

\subsubsection{Social Troubleshooting and Distributed Creativity}
Because biofoundries are expensive and complex, troubleshooting was often socially distributed. Introducing automation changes who is involved in the creative aspects of the work. Individuals now need a bigger and different set of social support. Participants noted the need for combined automation, programming and biology expertise to enable the creative exploration, while expertise is often distributed among different people.

Because biofoundry systems are complex and ``scary'' (P2, P4) to program, troubleshooting often becomes a collaborative effort between the scientist and core facility staff. Failures can be both monetarily expensive (``You basically have to pay somebody like [core staff name] to babysit this machine to make sure nobody breaks the million-dollar robot'' (P2)) and time-expensive (``If your culture goes bad... you've wasted two weeks'' (P8)). Core facility staff ``babysat'' (P2) robots to prevent costly damage, and experimental success, and thus scientific creativity, depended on the social dynamic and what one participant called an ``impedance match'' between biological intent and automation expertise (P3). In one case, close personal relationships enabled unusually deep collaborative troubleshooting, underscoring the social nature of creative work. Often, this creates a shift in responsibility and agency between the scientist and the core staff.
\begin{myquote}
    ``Because the BioFoundry's dumb, right? It just sits there and moves liquid from one place to another. But it is a complicated machine. And what it does is it shifts a lot of the responsibility for getting the experiment to work from the researcher onto whoever's managing the biofoundry.'' \participant{P2}
\end{myquote}
In some cases, particularly for industry biofoundries, researchers view the biofoundry as a ``fast intern'' (P3) that they trust to follow instructions but not to catch logic mistakes. This also impacts what each participant perceive as should be their job as opposed to someone else's.

\subsubsection{Tradaeoffs in Simplicity, Flexibility and Opaqueness in Biofoundry Design Shape What's Possible}
\label{sec:simplicity}

Learning to program is challenging, and not a standard part of training for specialized biologists. How, then, should the software interfaces for biofoundries be designed? There is a significant tension between simplification to enable access, and flexibility to enable creativity and power. P3 noted the difficulty in learning programming to work directly with the machines:

\begin{myquote}
[Scientists] want to be able to work with the code, because they think that that will give them maximum control over whatever biofoundry or cloud lab software that they have, and they'll be able to do a lot more things, make minute tweak and stuff to their will. But then, in reality, that is a hard skill to commit to learning.
\participant{P3}
\end{myquote}

Instead, P3's biofoundry uses no-code interfaces with GUI interactions like dropdowns, and default parameter suggestions.  However, other participants consistently noted the tension of interfaces that simplified interaction at the expense of scientific flexibility.
\begin{myquote}
    ``Scientific instrumentation software has been trying very hard to reduce everything down to press the big green button and get the thing that you want. But the closer you get to that, the more scientific flexibility you remove... The scientific instruments are complicated, and there's a lot of variability that they enable. 
    Simplicity is sort of antithetical to flexibility here. '' \participant{P5}
\end{myquote}

Other participants also noted that highly abstracted visual programming environments could paradoxically reduce agency, making troubleshooting harder because they hide ``what the system is actually doing'' (P8) and the underlying assumptions required to operate the robot safely. 

While automation enhances scientists' design analysis skills, it may also impose a ``bottleneck'' (P8) on creative learning if the system is too opaque. Errors could be caused by an untuned parameter buried deep in the software that the scientist did not know about and may struggle to find (P4, P8).

\section{Discussion}
\label{sec:discussion}





Automation is changing biology. 
In this moment, understanding how scientific creativity changes with automation can guide us to develop interfaces, social infrastructure, and automation paradigms that support and grow scientists' creativity.  
While a common narrative around biofoundries is that they free scientists from unnecessary manual labor to focus on the real, intellectual work of science, our interviews showed creative contributions remain deeply entwined with the material considerations of the biofoundry. 
\revised{\textbf{How, then, do we frame what biofoundries are, and envision what they should become?} 
In this section, we explore how framing biofoundries as \textit{creativity support tools} can help us better understand and design for changes in where creativity happens, how it is enacted, and what forms of reasoning it depends on.
}



\subsection{Biofoundries as CSTs}

Across our interviews, participants 
reached for metaphors to make sense of what biofoundries are and what kinds of work they enable. These metaphors matter: each foregrounds certain activities and obscures others, shaping expectations about authorship, responsibility, and creativity \cite{blackwell}. 



\revised{Framings of \textit{industrial infrastructure} (``cloud computing services'' (P6), a ``supercomputer'' (P4), or a ``factory'' (P3)) position biofoundries as centralized facilities run by experts, emphasizing standardization and separating intellectual design from experimental labor. Comparisons to \textit{advanced instruments} ``like a thermocycler'' (P7) or hopes that software be used ``like Photoshop'' (P3) suggest integration into everyday scientific workflows while preserving user agency. Meanwhile, metaphors like ``a very fast intern'' (P3) bring our focus to responsibility, capturing a system that executes instructions reliably but cannot account for human error.}

These metaphors share the focus on what work the biofoundry automates for you, from the entire execution flow (like in a factory) to a specific complex task (like a more capable thermocycler). This framing implicitly separates the scientists' creative work from the use of the biofoundry.  Creativity can be upstream from the factory (in the experimental design), or downstream (in the data interpretation), but the execution is bundled into a black box separate from the creative labor. 

In contrast, the framing of Creativity Support Tool emphasizes how tools help people explore problem spaces\revised{~\cite{Schon_1983}}, reflect on intermediate states\revised{~\cite{shneiderman}}, and iteratively refine ideas \revised{\cite{Schon_1983}}. 
Viewing biofoundries as CSTs foregrounds creativity that is already happening but often undervalued: translating protocols into executable workflows, anticipating failure modes, and designing experiments that remain interpretable under automation. Participants described pride not only in answering scientific questions, but in ``making a path'' (P6) for an experiment to work reliably on a robot. This resonates with prior work showing that creative contribution in technical domains often lies in structuring processes rather than producing final artifacts \cite{Suchman_1987, Schon_1983}.

Biofoundries offer a compelling yet underexplored site that extends CST theory into a domain with unusually high stakes. Biofoundries operate under conditions where failure can be expensive, irreversible, and biologically consequential. Materials are largely invisible, feedback is delayed, and much of the creative work occurs in maintenance and troubleshooting rather than artifact production.
In this context, creativity involves sustained interpretive labor that includes 
imagining failure modes in advance, reconstructing causal chains after the fact, and coordinating expertise across social and technical boundaries. 
This echoes accounts of creativity as an everyday, distributed practice \cite{Glaveanu_2014, Beuchley_2010}, thus extending CST theory into biology domains where creative work is inseparable from responsibility and risk.

\revised{In each of the following sections, we discuss how a lens from creativity support tools and related theories can help us understand our results and design for the biofoundry.} 

\subsection{\revised{Navigating Simplicity and Flexibility}}



\revised{
Our interviewees surfaced tensions around ``big green button'' interfaces (Sec. \ref{sec:simplicity}) that trade off simplicity and access with flexibility and power. While engineers may know the biofoundry's systems well, biologists may not know how to program; while academics might value understanding the full pipeline, scientists in industry might prefer to optimize for quick outputs. How do we navigate the tension between simplicity and flexibility in designing biofoundry interfaces?

One helpful framing is \textit{low floors, high ceilings, and wide walls}, drawn from Papert and Resnick's works on creative programming environments like Logo and Scratch \cite{Resnick_Robinson_2018}. While optimizing \textit{infrastructure} or an \textit{advanced instrument} might suggest the designer needs to find the right abstraction for the user's need, a CST suggests that the capabilities should both bring in novices (low floors) and support expert practice (high ceilings).  
Rather than a big green button for scientists or a raw programming interface for engineers, \textit{low floors and high ceilings} seek to scaffold developing expertise within the tools. }

\revised{
To navigate the differences between industry users, academics, and engineers, we can also adopt the idea of \textit{wide walls}. As a creativity support tool with wide walls, biofoundry software should allow many different interactions for many purposes. Rather than individualized interfaces for each user group, a system with wide walls acts as a flexible language for diverse expression.}

Participants' uncertainty about how much complexity to expose reflects that biofoundries are still being defined. Treating them as factories or services prematurely narrows their role, optimizing for throughput at the expense of exploration, learning, \revised{and the development of the user as a creative expert}. 
\revised{The design challenge, then, is not to build a ``big green button,'' but to create systems that lower barriers without flattening capability, support growth from novice to expert, and enable multiple ways of engaging with biological and computational work. In this view, biofoundries should be designed not as simplified services, but as full-spectrum creativity support systems.}

\subsection{\revised{Biofoundry as a Social and Distributed CST}}




Participants consistently described troubleshooting as a collaborative process where agency is distributed across scientists, automation engineers, core facility staff, and machines themselves. This reflects social and distributed models of creativity, emphasizing coordination across people, tools, and institutional roles \cite{Glaveanu_2014,Sawyer_2009}.
\revised{
No single actor has full visibility: scientists understand the biology, engineers understand the system, and logs provide only partial signals. Success depends on combining these perspectives.} 

Our findings discussed how creative outcomes in biofoundries depend on ``impedance matching'' (P3) between people with different forms of knowledge. Similar dynamics have been observed in fabrication labs and other AI-assisted creativity, where expertise is unevenly distributed \cite{kittur, Subbaraman_2025,Landwehr_2022}. What is distinctive in biofoundries is the asynmmetry of risk and access, as core staff often control the execution process, while researchers remain most credited for their scientific validity. 

\revised{
In distributed cognition, reasoning does not happen in any one individual, but emerges through coordination between partial representations distributed across people and artifacts \cite{cognition}. Scientists reason with biology, engineers reason with systems, and logs act as cognitive artifacts that store and transform partial representations. Diagnosing failure is therefore not an individual act, but a process of aligning these biological, computational, and operational layers.}

\revised{Role-based perspectives further reveal that distributed cognition in biofoundries is not only about coordination across people and artifacts, but about how different positions in the system define what counts as creative work. End-users emphasize intuition and experimental control, engineers emphasize system behavior and reliability, and founders emphasize abstraction and separation. These are not competing definitions, but partial views of a distributed cognitive system
.}

\revised{Sawyer and Dezutter's sociocultural view of creativity further reframes this process: creative outcomes emerge through interaction over time rather than from individual insight \cite{Sawyer_2009}. In this sense, the biofoundry is not just a tool, but a ``network of close colleagues'' \cite{Sawyer_2009} in which scientists, engineers, and machines collectively produce outcomes over time. 
The creative artifact is not just the biological result, but the workflow itself --- the constructed pathway that allows an experiment to succeed. Participants' descriptions of ``making a path'' (P6) for experiments to work highlight that the \textit{how} of automation is itself a site of creativity.
This helps explain why participants describe success as dependent on alignment, communication, and ``impedance matching'' (P3) across roles, and why authorship and ownership often feel blurred.
}

\revised{Taken together, these perspectives suggest that troubleshooting in biofoundries is 
a distributed cognitive process in which understanding emerges through coordination across people, tools, and time. Logs become cognitive artifacts, protocols become memory structures, and roles become functional components of reasoning.}

\revised{
Designing for biofoundries means 
supporting shared reasoning rather than isolated interaction, supporting alignment across roles, visibility into others' interpretations, and coordination across representations. In this view, biofoundries should be envisioned not just as platforms for executing experiments, but as systems for coordinating reasoning.}

\subsection{\revised{Embodied Reasoning in the Biofoundry}}

A key shift under automation is the transformation and loss of sensory engagement \cite{BainbridgeIroniesAutomation1983}. In bench biology, participants relied on embodied cues like smell, color, and texture to notice anomalies. In biofoundries, these ``soft senses'' are displaced by metadata traces, logs, and numerical summaries. \revised{This shift changes the fundamental feedback loop of experimentation. At the bench, action produces immediate sensory feedback, enabling rapid adjustment. In biofoundries, feedback is delayed, abstract, and symbolic.}

While similar shifts occur in digital fabrication, where designers lose tactile feedback but regain control through simulations, previews, and iterative material probing \cite{Vallgrarda_2007}, biofoundries differ in important ways. Digital fabrication workflows often emphasize fast, low-cost iterations, whereas biofoundry protocols are slow and expensive, with feedback arriving days or weeks later. As a result, participants invested creative effort in \textit{restoring visibility}, often through dry runs with food-coloring, parallel manual experiments, and comparative analysis across runs. 

\revised{Klemmer et al.'s account of embodied cognition \cite{Klemmer_2006} helps explain why this shift matters. Removing bodily interaction disrupts a core mechanism of reasoning. This is not simply a loss of data, but a change in the mode of cognition itself. Participants' sense of ``disconnectedness,'' difficulty interpreting logs, and reluctance to engage with abstract traces can be understood as the absence of embodied feedback loops that normally support intuition and hypothesis formation. 
}


Rather than encouraging playful tinkering, biofoundries demand what participants described as ``imagining everything that could go wrong before hitting submit.'' (P2) Participants mentioned how weeks of work could be wasted due to one uncaught mistake, and one interviewee even commented on how their biofoundry forbade anyone but the core staff to troubleshoot as mishandled errors could cause million-dollar machines to be out of service for the entire facility for weeks. This carefully anticipatory mode of creativity extends CST discussions beyond exploration towards foresight and resilience.

\revised{Participants' so-called ``workarounds'' (e.g., dry runs with food coloring, watching initial robot movements, iterative refinement etc.) can thus be understood not as inefficiencies, but as attempts to reconstruct embodied and situated feedback within a system that has removed it. These practices reintroduce perception-action coupling and create opportunities for mid-course sensemaking.}

\revised{This suggests a different design trajectory. Biofoundries should not be conceived as systems where scientists ``submit and walk away,'' but as environments that support ongoing interaction with experiments across time. Supporting perceptual coupling, enabling pauses and pivots, and translating abstract data into interpretable signals are not auxiliary features. They are central to reasoning.}




\revised{Taken together, these perspectives converge on the single shift that biofoundries are no longer simply tools that automate biology, but systems that reorganize creativity itself. If we take this seriously, then the question is no longer how to make biofoundries more efficient, but how to design them as systems that support distributed reasoning, embodied understanding, and sustained interaction. That is, how do we envision what biofoundries are and should be: not as black boxes, but as creativity support systems in the fullest sense.}

\subsection{Limitations and Future Work}

\revised{This study represents an initial foray into how creativity research can shape the future of biofoundries, a new and evolving space. While our recruitment sought to cover a wide variety of perspectives, this is not a comprehensive survey of the biofoundry community.} Our sample is primarily US-based. Significant innovation in biofoundries is also happening globally\footnote{https://www.biofoundries.org/}. 
We hope this work will encourage future research and design to focus on how biofoundries can support scientific creativity. 

\textit{How should software for biofoundries be designed?} Approaching the software interfaces for biofoundries as CSTs allows designers to foreground questions of reflection, exploration, ideation, and social and material contexts. In future work, we hope to design interfaces that elevate active problem solving and smoother transitions between manual bench work and automated execution. 

\textit{How should biofoundries incorporate AI tools?} This work suggests that clarifying where scientists situate creativity in their work can inform what types of AI intervention are appropriate or not, and how such interventions will affect agency, confidence, and problem solving across the DBTL cycle. Future work can dive deeper into how tools like protocol design assistants, automated problem identification, or automated data analysis can augment creativity, rather than remove it from the hands of scientists. 

\textit{How do we envision the future of biofoundries?}  As creativity support tools, biofoundries can not only increase scale, efficiency, and availability of biological experimentation, but also transform what creative activities we value, teach, and seek to support in science.

\section{Conclusion}

Biofoundries are often discussed as technical infrastructure for accelerating biological research. Our findings suggest a different emphasis, by framing biofoundries as CSTs. We emphasize how they reorganize the everyday practices through which scientific creativity is enacted. Rather than removing creativity from the laboratory, automation relocates it into cycles of designing, translating, troubleshooting, and learning.

\revised{While our findings are based on a limited heterogeneous sample set, they point to a broader challenge. As scientific work becomes increasingly automated, creativity becomes harder to identify, harder to support, and more dependent on how systems are designed. Re-envisioning biofoundries in this way offers a path forward.}
Through nine in-depth interviews with biofoundry experts, creativity emerged as embedded in the practical work of making experiments runnable on automated systems: anticipating failures before execution, adapting protocols to robotic constraints, reconstructing material outcomes from partial and delayed signals, and collaborating with facility staff. These activities were frequently framed as routine or supportive, yet they required judgment, improvisation, and sustained engagement with uncertainty.

Biofoundries also introduce distinctive creative conditions. Materials are largely invisible during execution, iteration is slow and costly, and agency is distributed across people, software, and machines. To recover opportunities for learning and sensemaking, scientists develop practices to restore visibility and control, including dry runs, parallel experiments, and close coordination with automation engineers. These practices highlight the unique constraints of biological experimentation at scale. We argue that viewing biofoundries as Creativity Support Tools foregrounds these dynamics and helps orient future designs toward systems that support distributed reasoning, embodied understanding, and sustained interaction.

\begin{acks}

\revised{This work was supported by the iBioFoundry: A research center supported by U.S. National Science Foundation (NSF) under grant No. [2400058]. We would like to thank all the interview participants for their time and insights. We are grateful to Huimin Zhao, George Heintz, Claudia Lutz, Nilmani Singh, Stephan Lane, Junyu Chen, Jingxia Lu, David Bianchi and members of the UIUC iBioFoundry for their support. We also thank members of the PICL lab for their valuable feedback.}

\end{acks}

\bibliographystyle{ACM-Reference-Format}
\bibliography{biofoundry}


\end{document}